\documentclass[aps,a4paper,superscriptaddress,twocolumn]{revtex4}
\usepackage{tensor}
\usepackage{graphicx}
\graphicspath{ {images/} }
\usepackage[utf8]{inputenc}
\usepackage{amsmath}
\usepackage{amssymb}
\usepackage{enumerate}
\usepackage{subfigure}
\usepackage{tabularx}
\usepackage{lipsum} 
\usepackage{float} 
\usepackage[colorlinks=true, pdfstartview=FitV, linkcolor=blue, citecolor=red, urlcolor=black, breaklinks=true]{hyperref}
\newcommand{\be}{\begin{equation}}
\newcommand{\ee}{\end{equation}}
\newcommand{\ben}{\begin{eqnarray}}
\newcommand{\een}{\end{eqnarray}}
\newcommand{\bes}{\begin{subequations}}
	\newcommand{\ees}{\end{subequations}}
\def\bal#1\eal{\begin{align}#1\end{align}}

\newcommand{\LL}{{\mathcal L}}

\newcommand{\pu}{\mathrm{\partial_{\mu}}}

\newcommand{\Pu}{\mathrm{\partial^{\mu}}}

\newcommand{\qt}[1]{``#1''}

\newcommand{\deriv}[2]{\ensuremath{\frac{d#1}{d#2}}}
\newcommand{\pb}[1]{\ensuremath{\partial_{#1}}}
\newcommand{\pr}{\ensuremath{\partial_{r}}}
\newcommand{\intg}{\ensuremath{\int_{\Sigma}dr\gamma}}

\newcommand{\ts}{\ensuremath{\tilde{\sigma}}}

\begin{document}
	\title{Impurity-doped stable domain walls in spherically symmetric spacetimes}
	
	\author{D. Bazeia}\affiliation{Departamento de F\'\i sica, Universidade Federal da Para\'\i ba, 58051-970 Jo\~ao Pessoa, PB, Brazil}
	\author{M. A. Liao}\affiliation{Departamento de F\'\i sica, Universidade Federal da Para\'\i ba, 58051-970 Jo\~ao Pessoa, PB, Brazil}
	\author{M. A. Marques}\affiliation{Departamento de Biotecnologia, Universidade Federal da Para\'iba, 58051-900 Jo\~ao Pessoa, PB, Brazil}
	
	\begin{abstract}
		In this work, radially symmetric domain wall solutions in the presence of impurities are investigated for both flat and curved $D+1$ spacetimes, with geometry generated by a rotationally invariant background metric. We have examined the constraints placed upon the model by Derrick's theorem, and found out the Bogomol'nyi bound and equations of the symmetric restriction to this theory. Impurity-doped versions of a $\phi^4$ model in two dimensions and a model with logarithmic potential in a Schwarzschild background have been explicitly worked out. The resulting configurations have been compared with those found in the homogeneous version of the theory, so that the effect of impurities in the form of solutions may be better appreciated. We have also generalized to higher dimensions some of the results that had been presented in the recent literature. These results relate to the possibility of BPS-preserving impurities, which we have found to still exist in the spacetimes considered in this work. We also investigate ways in which these results may be extended in a curved background.
	\end{abstract}
	
	\maketitle

\section{Introduction}\label{Intro}

The presence of topological structures is of great importance in high energy physics, as reported in Refs. \cite{B1,B2,manton} and in references therein. The standard and perhaps most known of such configurations are kinks, vortices and magnetic monopoles. Kinks appear in $1+1$ spacetime dimensions as real scalar field configurations, but vortices and monopoles require two and three spatial dimensions, as well as the addition of Abelian and non Abelian gauge fields, respectively. The study of these localized structures in the presence of impurities is also of current interest, as one can see from Refs. \cite{Benincasa,Hook,Tong,HY,AdamI,AdamII,BLMPLB,impu1,Cockburn,Ashcroft,Gud,BLM22,Wei}, to quote some recent investigations in the subject.

The inclusion of impurities has several distinct motivations, one of them being the possibility of modelling more realistic scenarios, and study how they may affect the otherwise standard results. In $1+1$ spacetime dimensions, the interesting spectral wall phenomenon was recently reported in kink-like collisions in the presence of an impurity \cite{impu1}. Also, in $2+1$ dimensions, fermionic impurities in  supersymmetric Chern-Simons theories were considered in \cite{Benincasa}. Moreover, in the study of vortices, the impurity can be of the electric or magnetic type, and may add different contributions; see, e.g, Ref. \cite{Tong}. The addition of impurities may also affect the scattering of vortices \cite{Ashcroft}.  Another motivation is related to integrability, with the integrable vortex equations being generalized to include magnetic impurities as well \cite{Gud}.

In this work, we focus on the study of topological structures in models in which a real scalar field is coupled to an impurity function in $D+1$ spacetime dimensions with a radially symmetric background metric.  We find a Bogomol'nyi bound and the respective first-order equations that must be solved by configurations which saturate this bound. We solve these equations and explore the modifications engendered in the system by the introduction of those inhomogeneities. The behavior of  solutions in the presence of localized impurities is investigated in two distinct scenarios, one in flat space and the other in the outer region of a Schwarschild black hole. We also generalize some of the results that were first presented in Refs.~\cite{AdamI, AdamII} for the $D=1$ case, and explore possible extensions of this generalization.

We organize the investigation as follows: in Sec. \ref{II} we present a general discussion, introducing the model and collecting the preliminary results. We then move on and elaborate on the Bogomol'nyi procedure and the presence of zero modes in Sec. \ref{III}. We illustrate the main results in Sec. \ref{IV}, considering two distinct scenarios, the flat or Minkowski and the Schwarzschild geometries. We go further on and discuss the case of form-preserving impurities in Sec. \ref{V}. We close the work in Sec. \ref{IV}, where we summarize the results and comment on some perspectives of future investigations.

\section{General discussion}\label{II}
Consider a real scalar field theory in $D+1$ spacetime dimensions. The standard action for such a theory can be obtained through integration of a Lagrangian density of the usual form $\LL_0=\frac{1}{2}\pu\phi\Pu\phi - U$, where $\phi$ is a real-valued function of spacetime coordinates and $U$ is a nonnegative potential with degenerate minima, which may, in general depend on both $\phi$ and the coordinates. The last condition on the potential is taken to allow for the possibility of topologically stable configurations.  In order to model the presence of impurities in our system, we must \textcolor{red}{add} an impurity function $\sigma(\mathbf{x})$, which may be coupled to the real scalar field through the inclusion of an additive term of the form $\LL_{\sigma}=-f\sigma$ to the Lagrangian, where $f=f(\mathbf{x}, \phi,\pu\phi)$ is a function that controls the coupling between $\sigma$ and $\phi$. Thus, we work with the action
\begin{equation}\label{S}
S\!=\!\! \int\! d^4x\sqrt{|g|}\left\{\frac12{\pu\phi\partial^{\mu}\phi} -f\sigma(\mathbf{x}) -U- \frac{\sigma^2(\mathbf{x})}{2}\right\},
	\end{equation}
where the last additive term was included for convenience and does not affect the equations of motion. If $\sigma$ is square integrable in the space coordinates, this addition amounts to a constant term $E_0=\int(\sigma^2/2)d^Dx$ on the energy. Else, this term  has no meaning of its own, but the full action~\eqref{S} may still make sense. We may alternatively interpret $U + \sigma/2$ as a deformation from the potential defined in $\LL_0$. 

The determinant $\sqrt{|g|}$ in~\eqref{S} represents a static background geometry. We assume spherical symmetry for the metric tensor $g_{\mu\nu}$, which is implicitly defined through the line element
\begin{equation}\label{metric}
ds^2=A^2(r)dt^2 - \left[B^2(r)(dr)^2 + \rho^2(r)d\Omega^2\right],
\end{equation}  
where $A, B$ and $\rho$ are smooth nonnegative functions of the radial coordinate and $d\Omega$ is a differential spawned by the $D-1$ angular variables of the model. Although the letter $r$ is used, we also include the case $D=1$, where no angular coordinates are present, even though the letter $x$ is more traditionally used in that case. In this work, we assume that the spacetime geometry is derived from a fixed background, so that Einstein's equations for the metric are not considered. The factor $\sqrt{|g|}$ may be derived from this metric, and calculation of this determinant leads to
\begin{equation}\label{jacobian}
\sqrt{|g|}=A(r)B(r)\rho(r)^{D-1}\omega(\theta_1,...\theta_{D-2})\equiv\gamma(r)\omega,
\end{equation} 
where $\omega$ generalizes the $\sin\theta$ factor known from spherical coordinates in the $D=3$ case. Note that polar coordinates in two spatial dimensions are naturally included as the case $\omega=1$. Although cylindrically symmetric geometries, which require the inclusion of a term of the form $\zeta^2(r)dz^2$ to $ds^2$, are not strictly included in~\eqref{metric}, the generalization of our results to this case is straightforward, and mostly amounts to a change in $\gamma(r)$. Thus, it shall suffice for our purposes to consider only geometries of the form~\eqref{metric}.

We shall now consider time-independent configurations. As was first proved by G.H. Derrick through the use of scaling arguments~\cite{Derrick}, stable solutions of this kind are not possible for a standard Lagrangian with self-interaction potential $U=U(\phi)$; see also Ref. \cite{Hobart}. Although stable time-dependent solutions might be found even within a standard model~\cite{stationary, Qballs}, a single-particle interpretation of the associated defects is difficult. Since Derrick's theorem is heavily dependent on the sign constraint found from requiring $\delta E$=0 under scaling, it may not hold in the presence of impurities. Although this possibility is of significant theoretical relevance, this work is most concerned with solutions that may be compared to the impurity-free case, defined by $\sigma=0$, and such a discussion is only meaningful for families of models that allow for stable topological defects even in the absence of impurities. For a discussion of defects in higher dimensions without this caveat, see~\cite{BLMPLB}. Thus, we shall here look for potentials that generate stable (or at least metastable) static configurations both in the presence and in the absence of impurities. In order to evade Derrick's theorem, one must change the energy functional, upon whose form the entire argument is constructed. In order to accomplish this, two approaches are possible:
\begin{enumerate}[(i)]
\item One may modify the matter Lagrangian density directly. In that direction, it was shown in Ref.~\cite{stable} that Derrick's argument may be evaded through the use of an $r$-dependent potential of the form $U=\tilde{U}(\phi)/r^{2D-2}$, with which configurations of the form $\phi=\phi(r)$ of the equations of motion have been found. Recently, Morris has expanded  this result to allow for more general rotationally-symmetric geometries.~\cite{Morris}. In our notation, this generalization amounts to substitution of the $r^{2-2D}$ multiplicative factor in the potential by $(B/\gamma)^2$.  In the same reference, it was shown that such a potential may arise naturally in the effective description of a real scalar field. This may be achieved, for example, if $\phi$ is non-minimally coupled to a Maxwell Lagrangian or to another scalar field solving an Euler-Lagrange equation of the form $\nabla_\mu\left[F(\phi)\nabla^{\mu}\chi\right]=0$, for some function $F(\phi)$. For details and other examples, see~\cite{Morris}. 	This approach has the advantage of working for any metric of the form~\eqref{metric}, including a Minkowski spacetime of any given dimension.
\item In rotationally symmetric curved spacetimes, one may choose the background geometry in such a way that the ensuing energy functional may be minimized when the Lagrangian is of canonical form. This approach was first used by Gonzales and Sudarsky~\cite{Gonzales} to find a stable, time-independent topological defect in a static Einstein universe, a result which has recently been extended~\cite{PRD101} to include a wider class of solutions in the same spacetime geometry. In Refs.~\cite{curved,stableDwall}, stable, spherical domain walls have been found for $(3+1)$ background metric tensor of the form $g_{\mu\nu}=\textrm{diag}(A^2(r), -A^{-2}(r),-r^2, -r^2\sin^2\theta)$. Ref.~\cite{stableDwall} in particular establishes precise criteria that must be satisfied by a metric of the aforementioned form in order to allow for stable domain walls. Unfortunately, these criteria include the condition that $A'(r)$ must have zeros in the region of interest, which exclude some important cases such as the Schwarschild, as well as the Reissner-N\"{o}rdstrom background, which had been ruled out in previous investigations~\cite{Palmer}. Despite those caveats, these results are still compatible with a large class of geometries, with the Schwarschild-Rindler AdS space being an important example~\cite{stableDwall}. 
\end{enumerate}

In both approaches, stable solutions have only been found under the assumption of some symmetry that allows the static field equations to be solved in terms of a single variable. Hence,  we shall henceforth assume rotational symmetry and thus take $\phi=\phi(t,r)$. Since the angular dependence of our problem is entirely contained in the determinant of the metric, the energy functional is proportional to $\Omega_D$, where $\Omega_D$ is a factor which depends on the spatial dimension considered and is found through integration of the volume form $\omega d{\theta_1}...d{\theta_{D-2}}$, with $\omega$ specified by~\eqref{metric} and~\eqref{jacobian}. By the principle of symmetric criticality~\cite{manton}, the equations of motion with the assumption of radial symmetry may be obtained as stationary points of the spherically symmetric Lagrangian
\begin{equation}\label{L}
\begin{split}
L =& \ \Omega_D\left\{\int_{\Sigma}dr\gamma\frac{\dot{\phi}^2}{2A^2(r)}  -\int_{\Sigma}dr\gamma \left[\frac{1}{2}\left(\frac{\pb{r}\phi}{B(r)}\right)^2 +U \right]\right \}\\
&-\Omega_D\intg \left(f\sigma +\frac{\sigma(r)^2}{2} \right),
\end{split}
\end{equation}
 where $\Sigma$ is the domain of integration in the radial coordinate and the dot denotes differentiation with respect to time. The field equations these configurations must satisfy are
\begin{equation}
\pb{t}\partial^{t}\phi- \frac{1}{\gamma}\pb{r}\left[\gamma\left(f_{\phi'}\sigma - \partial^r\phi\right)\right]+f_{\phi}\sigma+U_{\phi}=0,
\end{equation}
where the subscripts $\phi$ and $\phi'$  denote differentiation with respect the field and to $\phi'\equiv\pr\phi$, respectively.

In an infinite space, the domain of integration is given by $\Sigma=(-\infty,\infty)$ if $D=1$ and $\Sigma=[r_0,\infty)$, for some $r_0$, if $D\geq 2$. More generally, we may demand that $\Sigma$ is a union of closed sets. This last possibility becomes specially important if the metric~\eqref{metric} gives rise to an event horizon. Since $\Omega_D$ has no effect on the variation of $E[\phi]$, the field equation in the static case correspond to stationary points of $\varepsilon[\phi, \pr\phi]\equiv \frac{E[\phi,\pr\phi]}{\Omega_D}$. For static configurations, this functional has the form
\begin{equation}\label{E}
\varepsilon=\int_{\Sigma}dr\gamma\left\{\frac{1}{2}\left(\frac{\pb{r}\phi}{B(r)}\right)^2 +U + f\sigma +\frac{\sigma(r)^2}{2} \right \}.
\end{equation}
 The Euler-Lagrange equations for these configurations are of the form $\frac{\delta\varepsilon}{\delta\phi}=0$, or
\begin{equation}\label{SO}
\frac{1}{\gamma}\pb{r}\left[\gamma\left(f_{\phi'}\sigma - \partial^r\phi\right)\right]-f_{\phi}\sigma-U_{\phi}=0.
\end{equation}

We may use~\eqref{E} to investigate scaling arguments following the method of Derrick~\cite{Derrick} and Hobart~\cite{Hobart}. We thus perform the transformation $\phi(r)\to\phi(\lambda r)\equiv \phi_{\lambda}$, which implies $\varepsilon\to \varepsilon_{\lambda}$, where 
\begin{equation}\label{Hobart}
\begin{split}
\varepsilon_{\lambda}=&\int_{\Sigma}dy\gamma(y/\lambda)\left[\frac{\lambda}{2}\left(\frac{\pb{r}\phi_{\lambda}}{B(y/\lambda)}\right)^2 +\frac{U(\phi_{\lambda},y/\lambda)}{\lambda} \right ]\\
 &+ \int_{\Sigma}dy\frac{\gamma(y/\lambda)}{\lambda}\left[ f(y/\lambda)\sigma(y/\lambda) +\frac{\sigma(y/\lambda)^2}{2}\right].
\end{split}
\end{equation}
To allow for stability under this transformation, we must impose the well known condition
\begin{equation}
\left. \deriv{\varepsilon_{\lambda}}{\lambda}\right|_{\lambda=1}=0,
\end{equation}
which leads to the restriction
\begin{equation}\label{Derrick}
 I_1+I_2=I_3+I_4,
\end{equation}
where $I_1$ has exactly the same form as the energy in~\eqref{E}, while
\begin{equation}
\begin{split}
I_2=&\int_{\Sigma} dr\left\{\frac{1}{2}\left(\frac{\pb{r}\phi}{B(r)}\right)^2 + U(\phi,r)  \right.\\& \ \ \ \ \ \ \ \ \ \ \ \  \left.  + f\sigma +\frac{\sigma(r)^2}{2} \right\} r\frac{d\gamma}{dr},
\end{split}
\end{equation}
\begin{equation}
\begin{split}
I_3=&\int_{\Sigma} dr\gamma\left\{\left(\frac{\pb{r}\phi}{B(r)}\right)^2 + \frac{(\pb{r}\phi)^2r}{B^3(r)}\deriv{B(r)}{r} \right. \\   
& \left.  \ \ \ \ \ \ \ \ \ \ - r\frac{\partial U(\phi,r)}{\partial r} -rf\frac{d\sigma}{dr}  \right\}.
\end{split}
\end{equation}
and
\begin{equation}
I_4=\int_{\Sigma} dr\gamma\left\{ \frac{\partial f}{\partial \phi'}\pb{r}\phi-r\frac{d\sigma}{dr} - r\frac{\partial f}{\partial r} \right\}\sigma
\end{equation}
Both the geometry and the presence of the impurity conspire to make this expression far more complex than its counterpart in the one-dimensional theory without impurities. This added complexity has as its direct consequence the weakening of Derrick's theorem. There are now many ways in which condition~\eqref{Derrick} can be satisfied, since there are several independent terms as well as a great deal of liberty in the choice of Lagrangians that are able to evade Derrick's argument. It is noteworthy that most terms in the above integrals have an explicit dependence on the impurity. This means that one who endeavors to use a canonical potential in their investigations will find a much easier path than the one encountered in the impurity-free theory, in which the allowed geometries are strongly constrained, and where stable domain walls have only been found as a result of intense, and mostly recent, labor~\cite{Gonzales, PRD101, curved, stableDwall}. Although we shall not, for reasons already discussed, pursue this line of investigation, it must be emphasized that the above result seems to indicate that, when impurities are allowed, there exists a wider array of geometries capable of supporting domain-wall solutions. This is a physically relevant observation which warrants future investigation. 

To investigate stability, one should also take the condition  $\left.\frac{d^2\varepsilon_{\lambda}}{d\lambda^2}\right|_{\lambda=1}>0$ into account. This second derivative leads to a rather complicated expression, because of the number of terms in~\eqref{Hobart}, but it suffices to say that the two extra terms that appear because of the presence of the impurity are more than enough to allow for stable configurations, even if the form of the potential is not constrained. Scaling arguments for $\sigma=0$ in radially symmetric geometries can be found in Ref.~\cite{Morris}, with which our results can be compared.

\section{Bogomol'nyi procedure and zero modes}\label{III}

In order to find a minimum of~\eqref{E}, we may use a Bogomol'nyi~\cite{bogo} procedure to derive an energy bound for symmetric configurations. This is an efficient way to search for stable solutions of the radially symmetric field equations without imposing strong restrictions in the metric. To do this, we shall assume a potential of the form
\begin{equation}\label{pot}
U(\phi, r)=\frac{1}{2}\left(W_{\phi}(\phi)\frac{B(r)}{\gamma}\right)^2,
\end{equation}
where $W_{\phi}$ is the derivative of an auxiliary function $W(\phi)$. To allow for a Bogomol'nyi procedure, the coupling function must satisfy the constraint $f=\sqrt{2U} - \frac{\pb{r}\phi}{B(r)}$, or
\begin{equation}\label{f}
f(r,\phi,\phi')=\frac{B(r)W_\phi}{\gamma} - \frac{\pb{r}\phi}{B(r)}.
\end{equation} 
By completing the square in the $\varepsilon$ functional~\eqref{E} with the above constraints, we then find
\begin{equation}\label{Bgm}
\begin{split}
\varepsilon&=\int_{\Sigma}dr\gamma\left\{\frac{1}{2B^2(r)}\left[\pb{r}\phi -B(r)\left(\sigma-\sqrt{2U}\right)\right]^2 \right\} \\
&+\int_{\Sigma}dr\gamma (\pb{r}\phi)\sqrt{\frac{2U}{B^2(r)}}\geq \int dW\equiv \Delta W,
\end{split}
\end{equation}
where use has been made, in the last line, of~\eqref{pot}.  Thus, the energy is subject to the bound 
\begin{equation}\label{bound}
E\geq\Omega_D\int_{\Sigma}dr(\pb{r}\phi)W_{\phi}=\Omega_D\Delta W,
\end{equation}
where $\Delta W\equiv W(\phi(r))|_{\partial{\Sigma}}$. Saturation of this bound occurs if and only if $\phi$ satisfies the Bogomol'nyi equation
\begin{equation}\label{FO}
\pb{r}\phi = \sigma(r) B(r) + \frac{W_{\phi}B^2(r)}{\gamma}.
\end{equation}
Since both $\sigma$ and $B$ are functions of the radial coordinate alone, we may as well define $\tilde{\sigma}(r)\equiv \sigma(r)/B(r)$ to write, more simply,   
\begin{equation}
\partial^r\phi +\ts(r) + \frac{W_{\phi}}{\gamma}=0,
\end{equation}
without explicit reference to $B(r)$. Since no additional requirement was made about $\sigma(r)$, one may alternatively write the action~\eqref{S} in terms of $\ts$, which could then be refereed to as the impurity itself, with coupling function $\tilde{f}=B(r)\sqrt{2U} - \partial_r\phi$. Using this first-order equation it is possible, although somewhat tedious, to show that condition~\eqref{Derrick} is identically satisfied by any solution of the first-order equations, as should be the case.

We may find the zero modes of Eq.~\eqref{FO} through linearization of this equation.  Solutions of the first-order equation belong to a one-parameter family of functions which form a moduli space. This space may be parametrized by the zero of a domain-wall, or by any other parameter that completely specifies a BPS solution, just as the position of a fixed point. Let $X$ denote this parameter, and write
\begin{equation}\label{deltaphi}
	\phi(r, \tilde{X})-\phi(r,X)=\delta X\psi(r),
\end{equation} 
where both $\delta X\equiv \tilde{X}-X$ and $\psi$ are assumed to be small, in  the sense that higher powers of these quantities may be neglected. We may use this definition together with Eq.~\eqref{FO} to find
\begin{equation}
\frac{d\psi}{dr}=W_{\phi\phi}\frac{B^2(r)}{\gamma}\implies \psi=C e^{\int W_{\phi\phi}\frac{B^2(r)}{\gamma}dr},  
\end{equation}
where $C$ is a constant of integration. Apart from the factor $B^2/\gamma$ which accounts for the change in geometry, this calculation is identical to the one made for the standard one-dimensional case, both with and without impurities~\cite{AdamII, manton}. The constant $C$ may be specified by using~\eqref{deltaphi} together with an specification of $X$. Let, for definiteness, $X$ denote a zero of the field and consider a transformation $X\to X+ dX$. Since both $X$ and $X + dX$ are zeroes, we can deduce, exactly as in the one-dimensional case,
\begin{equation}
 \left. \frac{\partial\phi}{\partial r}(r,X)\right|_{r=X}dr +  \left.\frac{\partial\phi}{\partial X'}(X,X')\right|_{X'=X}dX'=0
\end{equation}
and, since $dX=dr$, one is readily led to $ \frac{\partial\phi}{\partial r}=- \frac{\partial\phi}{\partial X}$ at those points. Using this result together with~\eqref{deltaphi}, we can find $\psi(r_0)=\left.-\frac{\partial\phi}{\partial r}\right|_{r_0}$ and thus we obtain, finally,
\begin{equation}
C=\frac{B^2(r_0)}{\gamma (r_0)}-B(r_0)\sigma(r_0).
\end{equation}

When $D=1$, this result agrees perfectly with those found in~\cite{AdamII}, provided that differences in the conventions adopted are taken into account. We see that the impurity enters $\phi$ only through $C$. Since, however this value changes with $r_0$, the effect of the impurity is felt in all points of the moduli space and must thus change the scattering properties of these solutions, as expected on physical grounds. Within this approximation, one may derive a scattering Lagrangian of the form $L_s=\Omega_D\dot{X}^2 M(X)/2$, with $M(X)=\int dr\gamma \psi $ which, apart from the unimportant multiplicative factor and the more relevant $\gamma$, in the definition of M, is  identical to the corresponding one-dimensional Lagrangian~\cite{AdamII}.

\section{Examples}\label{IV}

In this section, we solve the first-order equations to find BPS solutions in the presence of localized impurities for two important geometries.
\subsection{Flat spacetime}

 As a first example, let us a deal with a plane geometry in two spatial dimensions. In the impurity-free case, solutions saturating the Bogomol'nyi bound for this geometry have already been found~\cite{stable}, and may be readily compared with our results. To define the potential, we may choose the auxiliary function $W(\phi)$ in the form $W(\phi)=\phi - \frac{\phi^3}{3}$, so that $W_{\phi}=1-\phi^2$. This choice results in the well-known $\phi^4$ potential. This leads to the family of equations
 \begin{equation}\label{Models}
		\pb{r}\phi = \sigma(r) + \frac{1-\phi^2}{\gamma},
	\end{equation}
 which are of the Riccati form~\cite{Riccati}. One important property of the Riccati equation is the fact that it may be converted into a second order, linear equation. Here, this is achieved through the substitution 
	\begin{equation}
		\label{xi}\phi=\gamma\frac{\xi'}{\xi},
	\end{equation}
	 which leads to the second order equation
\begin{equation}\label{Riccati}
\xi''+\frac{\gamma'}{\gamma}\xi'-\left(\sigma + \frac{1}{\gamma}\right)\frac{\xi}{\gamma}=0.
\end{equation}
Upon solving the above linear equation, one may obtain $\phi$ algebraically by means of~\eqref{xi}. This correspondence may be a useful tool in the search for BPS solutions, since techniques for solving second-order linear equations are readily available. Moreover,~\eqref{Riccati} allows one to make use of the existence and uniqueness theorems that are available for equations of this kind. 

\begin{figure}[ht]
	\flushleft
	\includegraphics[width=8cm]{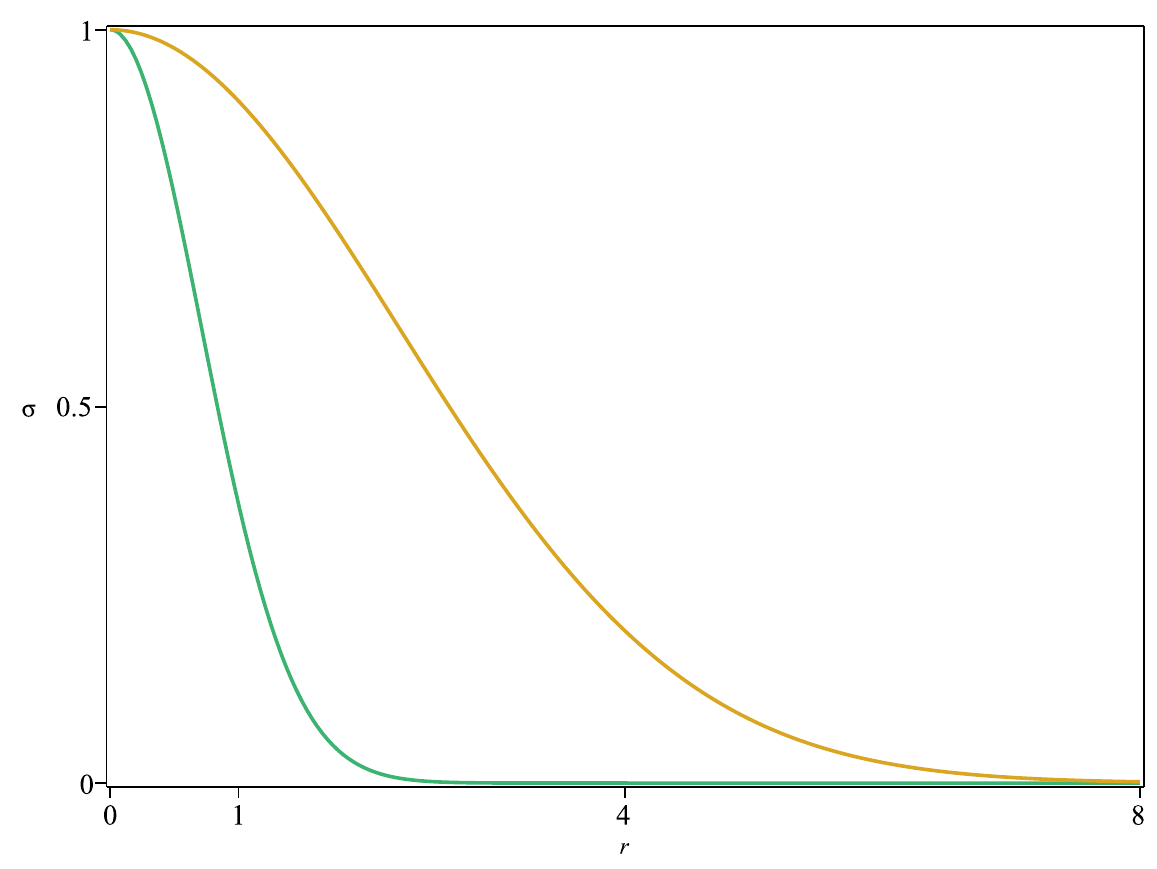}
	\caption{Impurity $\sigma(r)=\alpha e^{-\beta r^2}$ for $\beta=0.1$ (yellow) and $\beta=1$ (green). In both examples, $\alpha=1$}
	\label{fig1}
\end{figure}

\begin{figure}[ht]
	\flushleft
	\includegraphics[width=8cm]{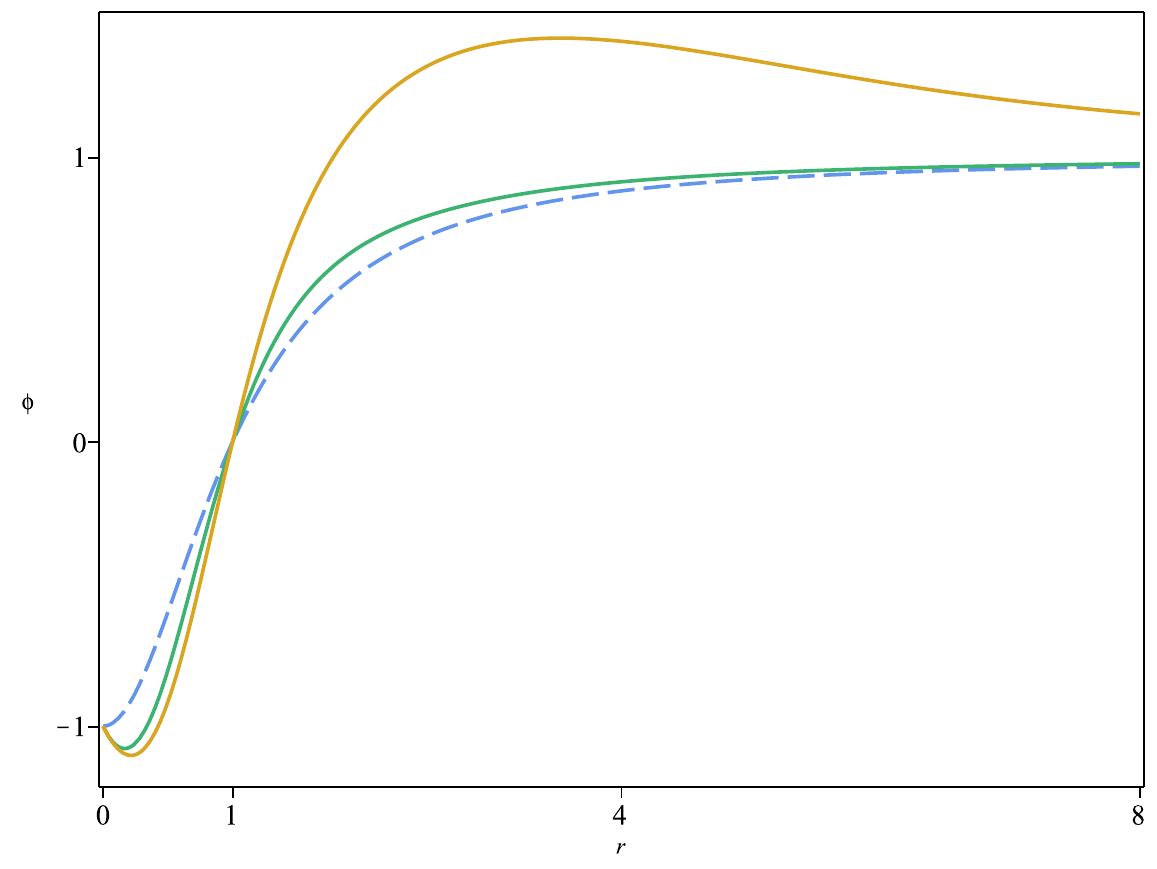}
	\caption{Solutions of~\eqref{Model1} with $W_{\phi}=1-\phi^2$ and impurity function given by~\eqref{sigma}, with $\alpha=1$ and $\beta=0.1$ (yellow), $\beta=1$ (green). The dashed blue line represents the impurity-free case, i.e., $\alpha=0$.}
	\label{fig2}
\end{figure}

In two-dimensional euclidean polar coordinates, the Bogomol'nyi equation~\eqref{Models} for a given sigma is
	\begin{equation}\label{Model1}
		\pb{r}\phi = \sigma(r) + \frac{1-\phi^2}{r}.
	\end{equation}
Let us exemplify the class of models defined by this choice of potential with a family of Gaussian impurities of the form
\begin{equation}\label{sigma}
\sigma(r)=\alpha e^{-\beta r^2},
\end{equation}
where $\alpha$ and $\beta>0$ are real constants. This impurity function can be seen in Fig.~\ref{fig1} for two values of the constant $\beta$, which controls how fast the impurity falls to infinity. The other constant, $\alpha$, may be important in scattering problems, but has not engendered any qualitative effect in the static solutions we are currently concerned with, so we fix it at unity. Impurities of this form were considered in~\cite{Cockburn, Ashcroft}, where they have been coupled to the topological vortices of Maxwell-Higgs theory. The solutions are determined up to a constant of integration $r_0$, which may be determined by specification of the zero of the scalar field. 
We have solved the above equation for the choices $\beta=0.1$ and $\beta=1$, and display the results in Fig.~\eqref{fig2}. For comparison, we also plotted the solution $\phi_0(r)=\frac{r^2-r_0^2}{r^2 + r_0^2}$, which corresponds to $\sigma=0$, in the same figure. One sees that the presence of an impurity changes the qualitative behavior of the solution, which is not a monotone function of the radial coordinate anymore, being initially a descending function of $r$, until it meets its minimal value (now not associated with a vacuum as before) and starts to grow. Indeed, the near-zero behavior of the solutions is $\phi\approx  Cr^2 -\alpha r $, where $C$ is a constant which can be determined numerically. For the solutions depicted in Fig.~\ref{fig2}, we have found $C=2.038$ ($\beta=0.1$) and $C=2.982$ ($\beta=1$). Since $C$ depends on $\beta$, both parameters of the impurity play a role in the determination of the position of this minimum of the solution. Another important difference from the impurity-free case is found in the emergence of a maximum for sufficiently small values of $\beta$. In the present example, this maximum ($\phi\approx 1.420$) is achieved for $r\approx 3.530$ when $\beta=0.1$.
\subsection{Schwarzschild background}
\begin{figure}[ht]
	\flushleft
	\includegraphics[width=8cm]{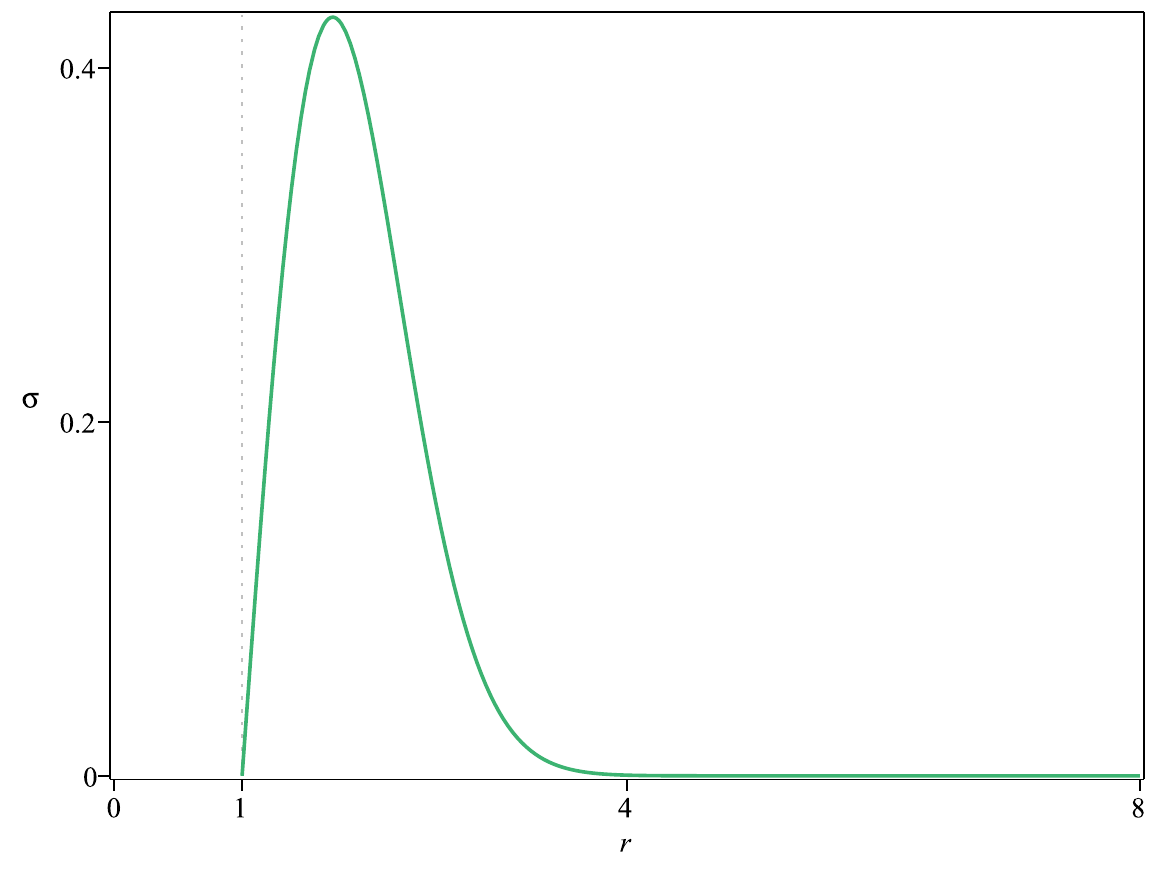}
	\caption{Impurity $\sigma(r)=(r-r_s)e^{-\left(\frac{r-r_s}{r_s}\right)^2}$, defined for $r>r_s$. Here, $r_s=1$.}
	\label{fig3}
\end{figure}

As a second example, let us consider a system interacting with the same family of impurities, but on a curved background in three spatial dimensions. Specifically, let us consider domain wall solutions on a Schwarzschild background, defined by the line element 
\begin{equation}\label{Sch}
\begin{split}
ds^2=&\left(1-\frac{r_s}{r}\right)dt^2 - \left(1-\frac{r_s}{r}\right)^{-1}dr^2\\ &+ r^2(r)d\Omega^2,
\end{split}
\end{equation}
where $d\Omega^2=d\theta^2+\sin^2\theta d\varphi^2$ and $r_s$ is the Schwarzschild radius, which defines the boundary of the event horizon of the black hole. Here, $A=B^{-1}=\sqrt{1-\frac{r_s}{r}}$. Instead of the $\phi^4$ potential of the previous example, we shall now choose a logarithmic potential of the form
\begin{equation}
U(r,\phi)=\frac{1}{2}\left(\frac{\phi \ln(\phi^2)}{r}\right)^2\left(1-\frac{r_s}{r}\right),
\end{equation} 
which presents minima at $\phi=0$ and $\phi=\pm 1$. We thus have $W_\phi=\phi\ln(\phi^2)$ and the first-order equations in the radially symmetric case must thus be of the form
\begin{equation}\label{FO2}
\pb{r}\phi = \frac{\sigma(r)}{\sqrt{1-\frac{r_s}{r}}} + \frac{\phi(r)\ln(\phi(r)^2)}{r^2}\left(1-\frac{r_s}{r}\right)^{-1}.
\end{equation}
 When $\sigma=0$, the above equation can be solved to find a solution 
 \begin{equation} \label{phi0}
\phi_0(r)=-e^{-\kappa\,\left(\frac{r}{r-r_s}\right)^{-2/r_s}},
 \end{equation}
where $\kappa$ is an arbitrary real constant. It is important to note that, although the function  above does solve the first order equation $\pb{r}\phi=\frac{\phi(r)}{r^2}\ln(\phi(r)^2)\left(1-\frac{r_s}{r}\right)^{-1}$ and the corresponding second-order equation, it does not behave in the way expected from a topological defect, because its asymptotic value does not belong to the vaccuum manifold of the theory. Indeed, for any finite value of $\kappa$, one can readily see from the analytical expression above that $\phi_0$ tends to a finite value that is completely defined by the constant $\kappa$. Such a solution still possesses a finite (though $\kappa$ dependent) energy due to the $r^{-2}$ factor in the potential, which ensures that the energy density goes to zero asymptotically. However, such a solution would only be possible if one imposes the somewhat arbitrary boundary condition $\lim_{r\to\infty}\phi_0(r)=-e^{-\kappa}$, which is difficult to justify on physical grounds. Let us see how this situation changes in the presence of an impurity. We choose a $\sigma$ function of the form

	\begin{equation}\label{sigma2}
 	   \sigma(r)=(r-r_s)e^{-\left(\frac{r-r_s}{r_s}\right)^2} \ \ \ \ (r>r_s).
	\end{equation} 
This impurity (Fig.~\ref{fig3}) exists only in the outer region of the Schwarzschild black hole, and it is zero at the horizon itself.  In our examples, we have chosen a Schwarzschild radius equal to unity. 
 
\begin{figure}[ht]
 	\flushleft
 \includegraphics[width=8cm]{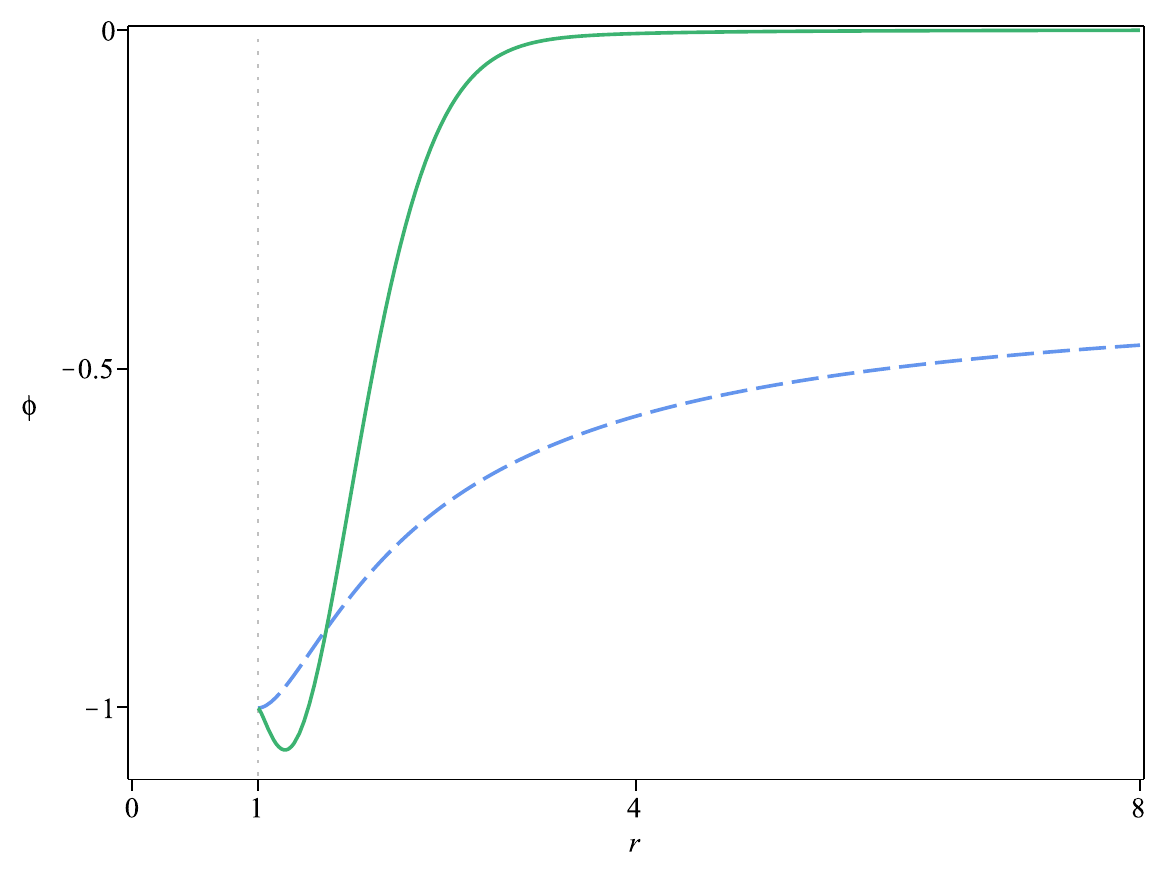}
 \caption{Solutions of \eqref{FO2} for $\sigma=0$, $\kappa=1$ (dashed blue line) and for $\sigma$ given by \eqref{sigma2} (green, solid line). Here, $r_s=1$.}
 	\label{fig4}
 \end{figure}

As in the previous example, the solution of~\eqref{FO2} cannot be written in closed-form, but we have calculated it numerically, and the result is depicted in Fig.~\ref{fig4} alongside the previously discussed solution~\eqref{phi0} of the impurity-free case. Unlike the latter, the impurity-doped solution \textit{does} reach the desired asymptotic value,  connecting two vacua of the model as expected from a topological solution, thus circumventing the problem  found in the impurity-free scenario. As in the previous example, the field is not a monotonic function of $r$ when an impurity is present, instead displaying a minimum at $r\approx 1.215$. Near $r=1$, the solution is of the form $\phi(r)\approx -1 - 2C_1(r-1)^{2}+ (r-1)^{3/2}$, where we have found an approximate value of $3.296$ for $C_1$. The last term, which only appears because of impurity, forces the field to decrease initially.

\section{Form-preserving impurities}\label{V}

A very interesting property found in~\cite{AdamII} for the one-dimensional case is the possibility of preserving the form of the kink when certain impurities are added to the system. One may wonder if such a possibility is also present in higher dimensions. Indeed, the second-order equation of the impurity-free case will be satisfied if and only if the two $\sigma$ terms in Eq.~\eqref{SO} cancel each other out exactly. This is possible if the impurity function solves the differential equation

\begin{equation}\label{Preserving}
\frac{1}{\gamma}\pb{r}\left[\frac{\gamma\sigma}{B(r)}\right] + \frac{W_{\phi\phi}B(r)}{\gamma}\sigma=0.
\end{equation}
 This condition generalizes the one found in Refs.~\cite{AdamI, AdamII}, with which it agrees when $D=1$. If the solution solves the first-order equation $\pb{r}\phi=W_{\phi}\frac{B^2(r)}{\gamma}$, corresponding to domain wall solutions, one may solve~\eqref{Preserving} to find
 \begin{equation}
\sigma(r)=\alpha\frac{B^2(r)}{\gamma^2\pb{r}\phi},
 \end{equation}
 where $\alpha$ is a real constant. For example, the solution \begin{equation}\phi_0(r)=\frac{r^2-r_0^2}{r^2 + r_0^2}\end{equation} of the $\phi^4$ model is preserved if any impurity of the family  \begin{equation}
\sigma_{\alpha}(r)=\frac{\alpha\left(r^2+r_0^2\right)^2}{r^4r_0^2}
 \end{equation} is added to the system. As expected, the configuration is no longer a BPS solution in the impurity-doped system, but it still solves the radially symmetric Euler-Lagrange equation of the system, as one may verify directly. The parameter $\alpha$ changes the shape of the impurity, making it thicker or thinner, and is important in scattering calculations~\cite{AdamII}. On the other hand, the remaining BPS equation of the impurity-free case, $\pb{r}\phi=-W_{\phi}\frac{B^2(r)}{\gamma}$, gives the result
 \begin{equation}
\sigma(r)=\beta\frac{\pb{r}\phi}{B(r)},
 \end{equation}
 where $\beta$ is also a real constant. The above equation preserves the \qt{antikink} solution of the impurity-free case.
 In two spacetime dimensions, the solution preserved through the use of~\eqref{Preserving} is not BPS~\cite{AdamII}, as it satisfies the field equation of the model, but cannot solve~\eqref{FO}. In one spatial dimension, not much can be done to change this situation, but when $D\geq 2$, new degrees of freedom are added because of the possible presence of background curvature. This raises the possibility of preserving a BPS solution in a Minkowski spacetime after \textit{both} an impurity and a nontrivial background metric are added to the system. In order to achieve this, the scalar field must solve the first-order equations in both systems, leading to the constraint equation
 \begin{equation}\label{constraint}
\frac{W_{\phi}}{\tilde{\gamma}}=\frac{W_{\phi}}{\gamma}B^2 + B\sigma,
 \end{equation}
 where $\tilde{\gamma}$ is a factor depending only on dimension and choice of coordinates, so that $\tilde{\gamma}=r$ in polar coordinates, $\tilde{\gamma}=r^2\sin\theta$ in spherical, etc. This constraint is satisfied provided $\sigma$ solves the algebraic equation
 \begin{equation}\label{sigmap}
\sigma= \frac{W_{\phi}}{B}\left(\frac{1}{\tilde{\gamma}}- \frac{B^2}{\gamma}\right).
 \end{equation}
 Thus, an impurity solving the above equation relates radial solutions which solve their respective BPS equations in different spacetimes. The requirement that the spacetimes be distinct is needed in order to prevent the vanishing of the $1/\tilde{\gamma} - B^2/\gamma$ factor in~\eqref{sigmap}. Inverting this logic of the above, one could also start with a fixed impurity function and solve Eq.~\eqref{constraint} for $B(r)$. The result is 
 \begin{equation}\label{B}
   B=\frac{-\sigma \pm \sqrt{\sigma^2-	\frac{4W_{\phi}^2}{\gamma\tilde{\gamma}}}}{2W_{\phi}}\gamma,
 \end{equation}
 which shows that the constraint equation can only be solved if $|\sigma|\geq \frac{2|W_{\phi}|}{\sqrt{\gamma\tilde{\gamma}}}$. The solutions in both spacetimes saturate the Bogomol'nyi bound~\eqref{bound}. Thus, we find
 \begin{equation}
\varepsilon=\int_{\tilde{\Sigma}}dr\tilde{\gamma}\frac{W_{\phi}}{\tilde{\gamma}}=\int_{{\Sigma}}dr{\gamma}\frac{W_{\phi}}{{\gamma}},
 \end{equation}
 so that the energy densities of the solutions satisfy $\rho=\frac{W_{\phi}}{\gamma}=\left(\deriv{\phi}{r}\right)^2\frac{\tilde{\gamma}}{\gamma}=\frac{\tilde{\gamma}}{\gamma}\tilde{\rho}$ and are thus related by the conformal factor $\frac{\tilde{\gamma}}{\gamma}$.
 
 \section{Discussion and final remarks}\label{VI}
 
 In this work, we have investigated domain wall solutions in real scalar field theories with impurities. Working in a $(D+1)$-dimensional spacetime with a spherically symmetric background metric, we have found first-order equations whose solutions possess the minimal energy compatible with radial symmetry, and presented some solutions of these equations, both in flat and Schwarzschild spacetimes. The ensuing defects have been compared to their impurity-free counterparts, allowing us to analyze the specific effects of the impurity doping on the behavior of the solutions. Next, we have investigated the possibility of doping the system with an impurity which preserves the form of an (anti)domain wall solution from an impurity-free theory. This investigation leads, in each case, to an equation that defines a family of impurity functions parametrized by a real constant, thus generalizing the results that were found in Ref.~\cite{AdamII} for the $D=1$ case. Finally, we have showed that, when $D\geq 2$, there exists an impurity which relates the \qt{standard} case investigated in~\cite{stable} to a different system in a curved spacetime. Both systems present identical solutions of their respective radial Bogomol'nyi equation. 
 
 An interesting perspective related to this last application lies in the generalization of the systems considered here to allow for coupling with gravity, in which case Einstein equations (or one of their generalizations) must be considered as well. When an impurity is added to the system, the energy-momentum tensor which acts as a source to these field equations is changed. One may thus wonder if it is possible that~\eqref{constraint}, which is a constraint equation in this work, may emerge naturally in a theory with gravitation. Other interesting paths for future investigations include the extension to theories with more than one real scalar fields, for instance, in the multi field scenario similar to the one considered in Ref. \cite{AdamX} in the one-dimensional case. Systems in which kink-like solutions are coupled to more complex defects, such as vortices, have been investigated in the absence of impurities (see, for example,~\cite{internal, multilayered} and references therein), and may be generalized to include impurities, in a way similar to the investigations  considered in \cite{Cockburn,Ashcroft}. Moreover, there are other aspects of the theory that merit further investigation, such as the search for time-dependent and non-BPS solutions, the study of defect scattering and formal proofs of existence and uniqueness for solutions, as well as the possibility of exploring other symmetries or even dropping the assumption of radial symmetry to further generalize our findings. Some of the above issues are presently under consideration, and we hope to return to them in the near future.

 \acknowledgements{The work is supported by the Brazilian agencies Coordena\c{c}\~ao de Aperfei\c{c}oamento de Pessoal de N\'ivel Superior (CAPES), grant No 88887.485504/2020-00 (MAL), Conselho Nacional de Desenvolvimento Cient\'ifico e Tecnol\'ogico (CNPq), grants No. 303469/2019-6 (DB) and 306151/2022-7 (MAM) and Paraiba State Research Foundation (FAPESQ-PB) grant No. 0015/2019. }




\end{document}